\documentclass[aps,prb,reprint,groupedaddress]{revtex4-1}
\usepackage{graphicx}
\usepackage{dcolumn}
\usepackage{bm}
\usepackage{epstopdf}
\usepackage{hyperref}
\usepackage[utf8]{inputenc}
\usepackage[T1]{fontenc}
\usepackage{lmodern}
\usepackage{amsmath}
\usepackage{amssymb}
\usepackage{booktabs}
\usepackage{color}
\usepackage{float}
\makeatletter
\newcommand*\nobreakhyphen{\hbox{-}\nobreak\hskip\z@skip}
\makeatother

\begin{document}


\title{Unconventional magnetic phase separation in $\gamma$-CoV$_2$O$_6$}


\author{L. Shen}
\affiliation{School of Physics and Astronomy, University of Birmingham, Birmingham B15 2TT, United Kingdom}

\author{M. Laver}
\affiliation{School of Metallurgy and Materials, University of Birmingham, Birmingham B15 2TT, United Kingdom}

\author{E. M. Forgan}
\affiliation{School of Physics and Astronomy, University of Birmingham, Birmingham B15 2TT, United Kingdom}

\author{E. Jellyman}
\affiliation{School of Physics and Astronomy, University of Birmingham, Birmingham B15 2TT, United Kingdom}

\author{E. Can\'evet}
\affiliation{Laboratory for Neutron Scattering and Imaging, Paul Scherrer Institut, CH-5232 Villigen PSI, Switzerland}

\author{J. Schefer}
\affiliation{Laboratory for Neutron Scattering and Imaging, Paul Scherrer Institut, CH-5232 Villigen PSI, Switzerland}

\author{Z. He}
\affiliation{State Key Laboratory of Structural Chemistry, Fujian Institute of Research on the Structure of Matter, Chinese Academy of Sciences, Fuzhou, Fujian 350002, China}

\author{M. Itoh}
\affiliation{Laboratory for Materials and Structures, Tokyo Institute of Technology, 4259 Nagatsuta, Midori, Yokohama 226-8503, Japan}

\author{E. Blackburn}
\affiliation{School of Physics and Astronomy, University of Birmingham, Birmingham B15 2TT, United Kingdom}



\begin{abstract}
We have explored the magnetism in the non-geometrically frustrated spin-chain system $\gamma$-CoV$_{2}$O$_{6}$ which possesses a complex magnetic exchange network.\ Our neutron diffraction patterns at low temperatures (\textit{T}\,$\leqslant$\,$T_{\mathrm{N}}$\,=\,6.6\,K) are best described by a model in which two magnetic phases coexist in a volume ratio 65(1)\,:\,35(1), with each phase consisting of a single spin modulation.\ This model fits previous studies and our observations better than the model proposed by Lenertz \textit{et al} in J.\,Phys.\,Chem.\,C 118,\,13981\,(2014), which consisted of one phase with two spin modulations. By decreasing the temperature from $T_{\mathrm{N}}$, the minority phase of our model undergoes an incommensurate-commensurate lock-in transition at $T^{*}$\,=\,5.6\,K. Based on these results, we propose that phase separation is an alternative approach for degeneracy-lifting in frustrated magnets.

\end{abstract}

\maketitle
\section{Introduction}
Magnetic frustration occurs when a system's total free energy cannot be minimized by optimizing the interaction energy between every pair of spins.\,This can be caused by competing interactions\,\cite{Seabra} or by geometry e.g.\ antiferromagnetic interactions on a triangular or tetrahedral unit\,\cite{Gardner}.\,As a result, the ground state of a frustrated magnet is often highly degenerate\,\cite{Diep}.\,The degeneracy can be lifted by perturbations such as additional interaction terms\,\cite{Gardner}, quantum fluctuations\,\cite{Savary}.\,Various exotic spin states may also result, as found by numerical simulations\,\cite{Okubo1, Okubo2}.\,Evidently, experiments are essential to verify the nature of the interactions, determine their parameters and to confirm the presence of any emergent states.

Quasi-one-dimensional (Q1D) spin-chain systems, wherein magnetic frustration often occurs, have attracted much attention due to their unconventional magnetic properties. For example, the Ising-like quantum ferromagnet CoNb$_{2}$O$_{6}$ (\textit{S}\ =\ 1/2) with an isosceles triangular spin lattice perpendicular to the chain direction (\textit{c}-axis) demonstrates a rich magnetic phase diagram in a transverse magnetic field (\textit{B}\,$\parallel$\,\textit{a}-axis)\,\cite{Lee}. More interestingly, its quasi-particle excitations near the paramagnetic quantum critical point reflect the E$_{8}$ symmetry that has been long predicted to exist in Ising chains\,\cite{Coldea}. 

Phase separation is a common phenomenon among colossal magneto-resistance (CMR) manganites and high-\textit{T}$_{c}$ superconductors\,\cite{Dagotto,EMERY}.\,It is usually a consequence of competing interactions.\,There are no constraints on the type of these interactions, though so far most phase separation phenomena require non-magnetic Hamiltonian terms (e.g.\,Coulomb interaction, spin-lattice coupling).\ Recently, phase separation possibly of purely magnetic origin was studied in SrCo$_{6}$O$_{11}$ where a `devil's staircase' is realised\,\cite{Matsuda},\,though the volume fractions of the competing phases were not determined.\,Dynamic phase separation has also been observed in the Q1D magnet Ca$_{3}$Co$_{2}$O$_{6}$\,\cite{Agrestini2} and possible microphases have also been reported therein\,\cite{kamiya, Prsa}. To our knowledge, static or dynamic phase separation exclusively caused by magnetic interactions on a non-geometrically frustrated lattice has not been observed until now.\par
\begin{figure*}
	\centering
	\includegraphics[width=0.99\textwidth]{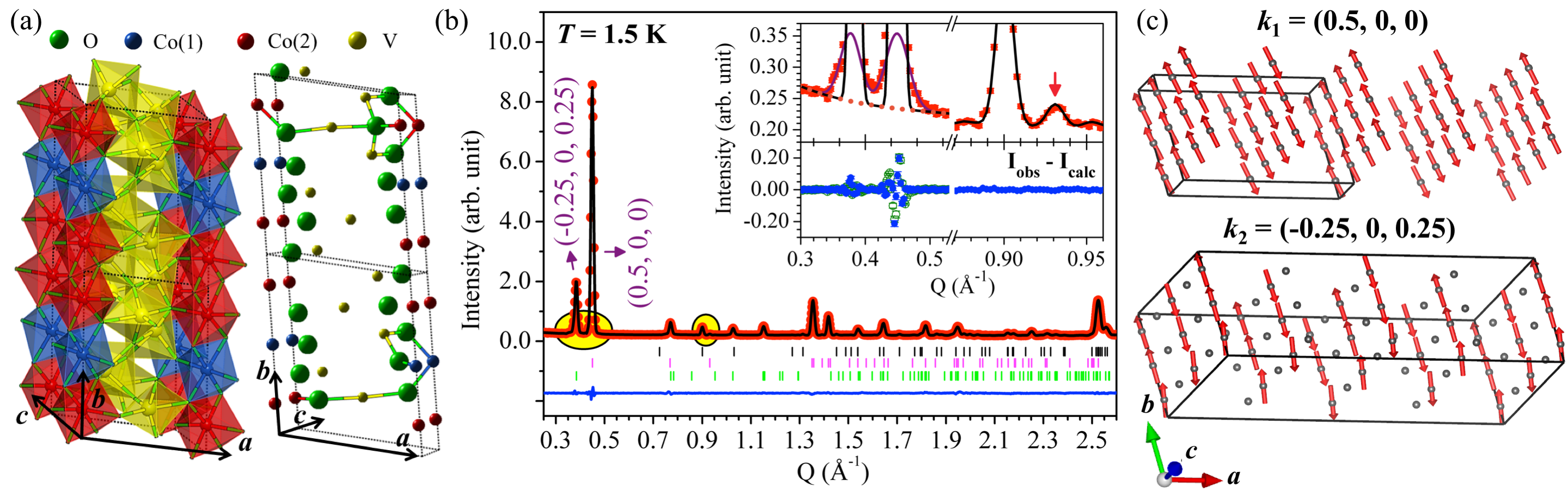}
	\caption{(a) Crystallographic structure of $\gamma$CVO. Oxygen anions (omitted for clarity) occupy the corner of the shaded polyhedra. Possible NN spin exchange paths for Co(1) and Co(2) are displayed separately in two unit cells. (b) Neutron powder diffraction pattern measured at $\lambda$\,=\,4.5\,\AA, \textit{T}\,=\,1.5\,K. The red solid dots are experimental observations.\,The black and blue lines are the calculated pattern and the difference using the \textit{2}-phase model.\,Black, pink and green vertical bars mark the nuclear, \textit{$k_{1}$}- and \textit{$k_{2}$}- modulated Bragg positions, respectively.\,Upper inset: Enlarged view of the shaded area in the main panel. The black solid line is the Rietveld refinement to the long-range order. The purple solid line is a Gaussian fit to the short-range correlations. The dotted line is a 3-degree polynomial fit to the background between 0.2 and 0.6\,\AA$^{-1}$. Lower inset: differences between the observed and calculated intensities in the low-\textit{Q} region based on the two background treatment methods described in the main text. (c) The spin structures, reproduced by VESTA\ \cite{VESTA}, in the $k_{1}$ (upper panel) and $k_{2}$ (lower panel) phases, respectively. The black frames display the size of the corresponding magnetic unit cells.}
	\label{fig:1}
\end{figure*}
We report magnetic phase separation in the triclinic cobaltate compound $\gamma$-CoV$_{2}$O$_{6}$ ($\gamma$CVO). $\gamma$CVO has space group \textit{P$\bar{1}$} with edge-sharing CoO$_{6}$-octahedra arranged in zigzag chains along the crystallographic \textit{b}-axis.\,These chains are well separated by a VO$_{4}$-VO$_{6}$ polyhedral-network between them (Fig.\,\ref{fig:1}a)\,\cite{MULLER}.\,Unlike its polymorph $\alpha$-CoV$_{2}$O$_{6}$ ($\alpha$CVO), the transverse nearest neighbour (NN) exchange in $\gamma$CVO must involve V$^{5+}$\,\cite{He}.\,This significantly weakens the interchain exchange interaction strength as evidenced by a lower ordering temperature in $\gamma$CVO\,\cite{He, He2, Kimber}.\,As shown in Fig.\,\ref{fig:1}a, there are two inequivalent cobalt sites, Co(1) and Co(2). For the Co(2)-Co(2) exchange, there is only one Co$^{2+}$-O$^{2-}$-V$^{5+}$-O$^{2-}$-Co$^{2+}$ (COVOC) path along the \textit{a}-axis (Fig.\,\ref{fig:1}a).\,In contrast, two very similar COVOC paths are found along the \textit{c}-axis, affording the possibility of the so-called `random frustration' caused by competing interactions\,\cite{Gardner}.\,For the Co(1)-Co(1) exchange, no NN COVOC path is found along the \textit{a}-axis and only one such path is located along the \textit{c}-axis. Surprisingly, a skew path between interchain Co(1) and Co(2) sites is also found. Its length is close to those of the transverse ones, meaning these skew paths are just as important for the magnetic structure. First of all, they can set up correlations between Co(1) spins along the \textit{a}-axis. Second, since the intrachain exchange is mainly ferromagnetic, an antiferromagnetic skew exchange would complicate the final magnetic structure or even lead to further frustration.
\section{Experimental methods}
Both single crystals and powders of $\gamma$CVO were synthesized for our investigations; details of the synthesis are given in Ref.~\onlinecite{He2}. The room-temperature nuclear structure of the powder sample was studied on a Siemens D5000 Powder Diffractometer (Cu K\,$\alpha$).\ All observed Bragg reflections can be well fitted by the triclinic lattice solution described in Ref.~\onlinecite{MULLER}; neither a second crystallographic phase nor structural disorder could be resolved (see the Supplemental Material). Magnetic susceptibility data were collected using a Magnetic Property Measurement System (MPMS, Quantum Design).\,The existence of magnetic frustration in $\gamma$CVO is experimentally supported by the commonly used frustration index \textit{f} = |\textit{$\theta$}$\rm{_{CW}}$/\textit{T}$\rm{_{N}}$| $= 1.66(3)$ (\textit{$\theta$}$\rm{_{CW}}$:\,Curie-Weiss temperature, \textit{T}$\rm{_{N}}$:\,N\'eel temperature)\,\cite{Gardner}.\,We carried out diffraction measurements on powder samples using the cold neutron powder diffractometer DMC at the Swiss Spallation Neutron Source (SINQ). Two neutron wavelengths, 2.4586\,\AA{} and 4.5\,\AA{}, were used.\,The longer wavelength provided the necessary angular resolution to distinguish the magnetic Bragg peaks.\ Six grams of powder was loaded into a thin Al cylinder (6\,mm in diameter) and then into a cryostat to probe temperatures down to 1.5\,K. The obtained diffraction patterns were refined using the Rietveld method in the FullProf package\,\cite{FullProf}. Single crystal neutron diffraction measurements were performed on the TriCS instrument at SINQ.\,These data (not shown here) confirm the propagation vector \textit{$k_{1}$}\,=\,(0.5,\,0,\,0) of the magnetic structure found by Kimber \textit{et al.}\,\cite{Kimber} and Lenertz \textit{et al.}\,\cite{Lenertz}, but we did not find peaks corresponding to the second propagation vector (0.25,\,0.5,\,0) proposed in Ref.~\onlinecite{Lenertz}.\,Furthermore, we find a magnetic Bragg peak at $Q$\,$\simeq$\,1.03\,\AA$^{-1}$ in our powder diffraction profiles (Fig.\,\ref{fig:1}b) that cannot be indexed using either of the previously found propagation vectors. 
\section{Results and discussions}
On cooling the system down to 1.5 K from the paramagnetic state, magnetic Bragg peaks are observed in the low-\textit{Q} region (Fig.\,\ref{fig:1}b).\ The refined lattice parameters\ (Table\,\ref{tab:I}) are consistent with previous works\,\cite{Kimber,Lenertz}.\,In addition to the \textit{$k_{1}$}\,=\,(0.5,\,0,\,0) wavevector proposed by Kimber \textit{et al.}\,\cite{Kimber}, corresponding to ferromagnetic \textit{bc}-planes antiferromagnetically coupled along the \textit{a}-axis, we find that a second propagation vector \hbox{\textit{$k_{2}$}\,=\,(-0.25,\,0,\,0.25)} is required to index the rest of the peaks.\,The in-plane spin modulations of $k_{1}$ and $k_{2}$ are shown schematically in Fig.\,\ref{fig:1}c.

We find short-range correlations down to the lowest temperature probed (1.5\,K).\,Their contributions below the incommensurate-commensurate lock-in transition $T^{*}$\,=\,5.6\,K were treated in two self-consistent ways. First, two Gaussian functions were used to fit the diffuse profiles on the tails of the main peaks at \textit{$Q_{1}$}\,=\,$(-0.25,\,0,\,0.25)$ at $\sim$ 0.39\,\AA $^{-1}$ and \textit{$Q_{2}$}\,=\,(0.5,\,0,\,0)\, at $\sim$ 0.45\,\AA $^{-1}$, respectively. The background in this region (0.2\ $\leqslant$\ \textit{Q}\ $\leqslant$\ 0.6\,\AA$^{-1}$) was fitted using a 3-degree polynomial function. These results are displayed in the upper inset of Fig.\,\ref{fig:1}b. The summed intensities of the Gaussian short-range order and polynomial background were loaded into a background file for the long-range order determination\,\cite{FullProf}; these background points were fixed in our Rietveld refinments. In the second approach, we did not treat the intensities arising from the short-range order separately. In other words, they were regarded as a part of the background. We constructed a background file which contains 24 points between 0.2 and 0.6\,\AA$^{-1}$ to cover the \textit{Q}-region `contaminated' by the short-range correlations; these points were refined using the Rietveld method. Concerning the background at high-\textit{Qs}, we constructed a background file and used it in both approaches; we did not refine these points since they are very flat (Fig.\,\ref{fig:1}b). As shown in the lower inset of Fig.\,\ref{fig:1}b, the two approaches described above produce almost identical residuals. We will further discuss the short-range correlations below.
\begin{table}[h!]
  \centering
  \caption{Magnetic and lattice parameters of $\gamma$CVO at $T\,=\,1.5$\,K. Constraints on the spin orientations for the $k_{2}$ modulation have been applied; see main text for details. $\rm{\overline{Co(2)}}$ is the central inversion replica of Co(2). The isotropic displacement parameters ($B_{iso}$) and V atomic positions were fixed to the values at 2\,K reported in Ref.~\onlinecite{Kimber}.\,Lattice parameters, O and Co positions were refined using data at $\lambda$ = 2.4586\,\AA{}. For the two phase scenario, three sets of Rietveld factors, corresponding to the minimal model\,($^{\bullet}$),\,inequivalent\,($^{\dagger}$) and equivalent\,($^{\ddagger}$) spin canting on Co(2)- and $\rm{\overline{Co(2)}}$- sites\,(see text), are listed.}
  \label{tab:I}
  \begin{ruledtabular}
  \begin{tabular}{c|ccc}
  Scenario I&&2-\textit{k}&\\
  \textit{a},\,\textit{b},\,\textit{c}\,(\AA)&7.1515(4)&8.8555(3)&4.7951(2)\\
    $\alpha$,\,$\beta$,\,$\gamma$\,($^{\circ}$)&90.144(5)&93.948(2)&102.110(6)\\
\hline
Moments&$M_{a}$ ($\mu_{B}$)&$M_{b}$ ($\mu_{B}$)&$M_{c}$ ($\mu_{B}$)\\
\hline
   Co(1)\,:\,$k_{1}$&-0.5(2)&2.5(1)& 0.3(3) \\
   Co(2)\,:\,$k_{1}$&0.2(1)&2.44(7)&-0.5(2)\\
   $\rm{\overline{Co(2)}}$\,:\,$k_{1}$&0.2(1)&2.44(7)&-0.5(2)\\
   Co(1)\,:\,$k_{2}$&-0.4(1)&2.0(6)&-0.01(1)\\
   Co(2)\,:\,$k_{2}$&-0.21(4)&1.0(2)&-0.003(4)\\
   $\rm{\overline{Co(2)}}$\,:\,$k_{2}$&-0.5(1)&2.5(5)&-0.01(1)\\
\specialrule{1.5pt}{1pt}{1pt}
 Scenario II$^{\dagger}$&&2-phase&\\
  \textit{a},\,\textit{b},\,\textit{c}\,(\AA)&7.1524(4)&8.8560(3)&4.7954(2)\\
    $\alpha$,\,$\beta$,\,$\gamma$\,($^{\circ}$)&90.137(6)&93.949(2)&102.122(7)\\
\hline
Moments&$M_{a}$ ($\mu_{B}$)&$M_{b}$ ($\mu_{B}$)&$M_{c}$ ($\mu_{B}$)\\
\hline
   Co(1)\,:\,$k_{1}$\,[65(1)\,\%]&-1.7(3)&2.9(3)&1.1(3)\\
   Co(2)\,:\,$k_{1}$[65(1)\,\%]&-1.1(2)&3.1(1)&-0.2(2)\\
   $\rm{\overline{Co(2)}}$\,:\,$k_{1}$[65(1)\,\%]&-1.1(2)&3.1(1)&-0.2(2)\\
   Co(1)\,:\,$k_{2}$[35(1)\,\%]&-0.69(4)&3.3(2)&0.008(4)\\
   Co(2)\,:\,$k_{2}$[35(1)\,\%]&-0.57(5)&2.8(2)&1.5(4)\\
   $\rm{\overline{Co(2)}}$\,:\,$k_{2}$[35(1)\,\%]&-0.65(2)&3.1(1)&-0.008(2)\\
 \specialrule{1.5pt}{1pt}{1pt}
 Rietveld factors&R$_{p}$ (\%)&R$_{wp}$(\%)&$\chi^2$\\
 \hline
 2-\textit{k}&6.29&5.78&4.796 \\
 2-phase$^{\bullet}$&6.25&5.77&4.749 \\
 2-phase$^{\dagger}$&6.20&5.72&4.657 \\
 2-phase$^{\ddagger}$&6.20&5.77&4.728 \\ 
  \end{tabular}
  \end{ruledtabular}
\end{table}

Although rare, multi-$k$ structures have been predicted and experimentally confirmed in some frustrated systems\,\cite{Stewart, Javanparast, Okubo1, Okubo2, Bourgeois, Pregelj}.\,We therefore propose two possible magnetic structures for $\gamma$CVO: (I)\,a single phase with 2-$k$-modulation, or (II)\,two 1-$k$ phases\,(phase separation). As shown by the Rietveld factors in Table\,\ref{tab:I}, both scenarios turn out to fit the data reasonably well, although with some caveats.\,Possible phase differences between the two inequivalent Co-sites and between the two modulations have been fixed to zero, since we found that these parameters either resulted in unphysically large magnetic moments or did not converge within the fitting resolution.\,We could not solve exactly the spin orientations modulated by $k_{2}$ in either scenario, since the relevant free parameters were highly correlated, resulting in unphysically large standard deviations in the Rietveld refinements.

We have also tested a `minimal model' for each scenario where all spins modulated by $k_2$ lie along the \textit{b}-direction; this is based on the assumption of Ising-like anisotropy along the crystallographic \textit{b}-axis\,\cite{He2,Lenertz,Drees}. This minimal model was then relaxed by allowing spin canting in the \textit{ab}-, or \textit{bc}- plane on each Co-site.\,For the 2-\textit{k} single phase scenario, this canting does not improve the original refinement produced by the minimal model, and so the corresponding spin orientations are fixed to the \textit{b}-axis.\,In a triclinic lattice, we note the spins will still have components in the \textit{ac}-plane even if the \textit{b}-axis Ising anisotropy is strictly followed (Table\,\ref{tab:I}). The refinement is not sensitive to additional spin canting on Co(1)-sites in the phase separation scenario (fixed along the \textit{b}-axis for these sites in Table\,\ref{tab:I}), but it is considerably improved by including canting in the \textit{bc}-plane on Co(2)-sites (see below).

Both scenario I and II fit the data reasonably well. However, the global average of the magnetic moment along the \textit{b}-axis ($\overline{M}_{b}$) obtained by the 2-\textit{k} solution is 4.3(3) $\mu_{B}$.\,This is close to the value in $\alpha$CVO where there is large spin-orbit coupling (SOC) \cite{He, Lenertz2, Markkula, Hollmann}.\,Crystallographic structure analysis shows that the distortion of the CoO$_{6}$-octahedron is much weaker in $\gamma$CVO than in $\alpha$CVO\,\cite{Wallington}.\,This leads to a very small orbital contribution to the total moment in $\gamma$CVO, as revealed by X-ray magnetic circular dichroism (XMCD) spectroscopy and theoretical calculations\,\cite{Hollmann,Kim}.\,The result is a global average spin moment of $\sim$\,3.2\,$\mu_{B}$/Co, mainly pointing along the \textit{b}-axis, in agreement with magnetization measurements\,\cite{Kimber,Drees,Lenertz,He2}. We point out that the 2-$\vec{k}$ solution is \emph{inconsistent} with this value.\ On the other hand, the phase separation model produces $\overline{M}_{b}$ = 3.04(9) $\mu_{B}$/Co, in excellent agreement with magnetization, XMCD data, as well as theoretical predictions\,\cite{Hollmann,Kim,Kimber,Drees,Lenertz,He2}. 

We will now discuss the magnetic structure of this phase separation scenario in detail.\ Previous susceptibility measurements on $\gamma$CVO single crystals\,\cite{He2} show that the Co ions still possess Ising-anisotropy along the crystallographic\ \textit{b}-axis.\ Recently, this anisotropy has been challenged by a time-of-flight inelastic neutron scattering study which suggests that one-dimensional magnetism along the \textit{b}-axis is \emph{not} sufficient to address all of their observations\,\cite{Wallington}.\ According to our refinement, the global average moment ($\overline{M}$) is 3.17(8) $\mu_{B}$/Co. When we compare this to $\overline{M}_{b}$ we see that bulk Ising-anisotropy is mostly maintained in $\gamma$CVO. On the other hand, we find that canting in the \textit{ac}-plane for spins in the $k_{1}$ phase is necessary to match some very weak reflections [e.g. (0.5, 1, 0) in Fig.\,\ref{fig:1}b(inset)].\,For example, the refined structure of the Co(1)-spins in the $k_{1}$ phase shows components along all 3 crystal axes (Table\,\ref{tab:I}). Since the projections of $M_{b}$ on both \textit{a-} and \textit{c-} axes are weak,\ e.g.\,-0.61\,$\mu_{B}$/Co(1) and -0.01\,$\mu_{B}$/Co(1),\ respectively,\ in the $k_{1}$ phase, the additional non-negligible \textit{in-plane} magnetic moments obtained in our refinements strongly indicate that the spins in $\gamma$CVO do not lie solely along the \textit{b}-axis. This might be related to the complex CoO$_{6}$-octahehral distortion seen in this compound\,\cite{Kim,Hollmann}. 

By relaxing from the `minimal model', we can estimate the strength of spin canting in the $k_{2}$ phase. By allowing canting in the \textit{bc}-plane on the Co(2)-sites, i.e.\,29(8)$^{\circ}$ towards the \textit{c}-axis, the refinement quality characterized by the three Rietveld factors is considerably improved (Table\,\ref{tab:I}). This canting angle changes to 19(9)$^{\circ}$ and the Rietveld factors are increased if we keep the inversion symmetry between Co(2)- and $\rm{\overline{Co(2)}}$- sites. These results support the breakdown of inversion symmetry on Co(2)-sites in the spin lattice. This breakdown is only allowed in the $k_{2}$ phase based on the representation analysis.
\begin{figure}
	\centering
	\includegraphics[width=0.45\textwidth]{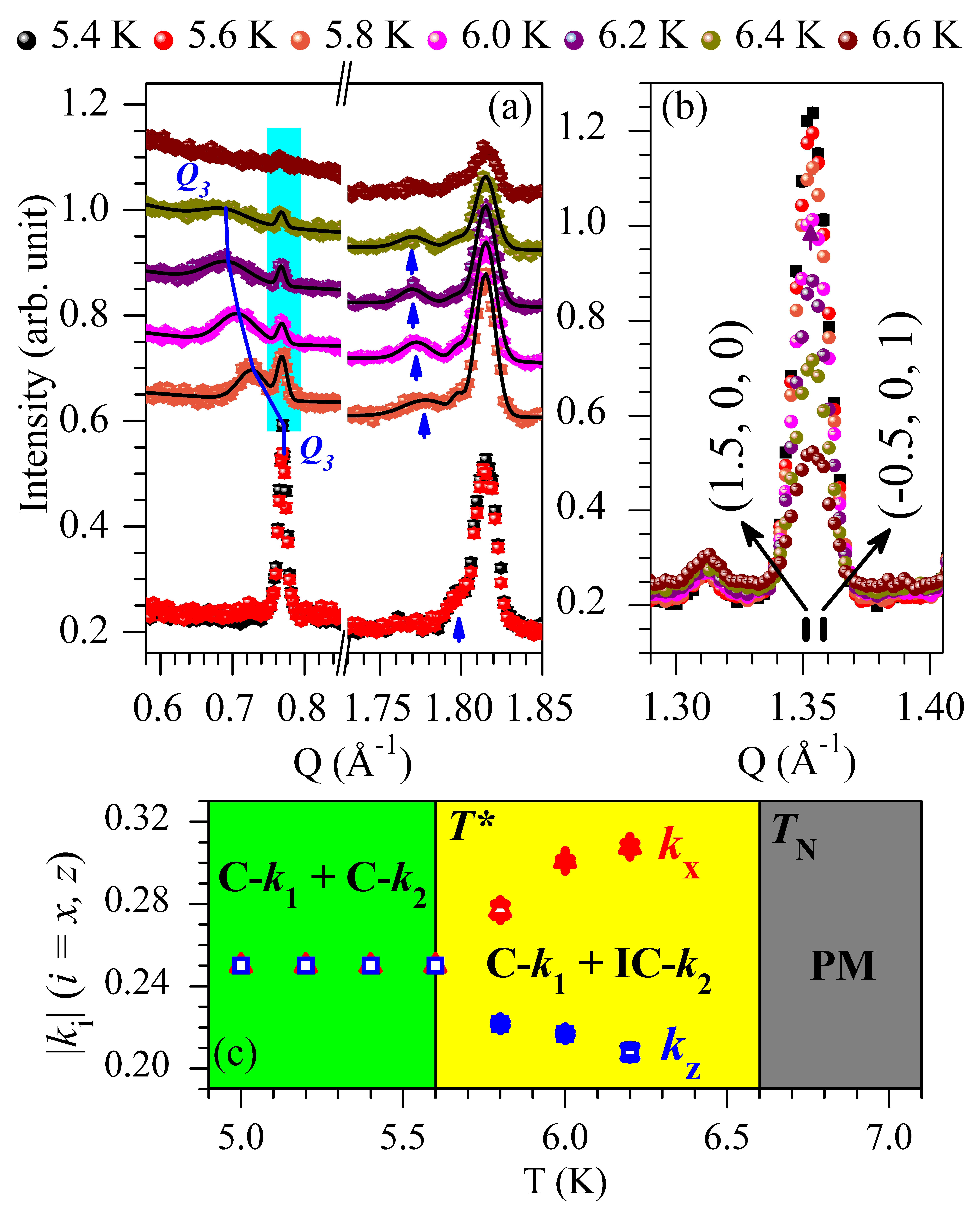}
	\caption{(a) Selected regions of the powder diffraction patterns between 5.4\,K and 6.6\,K. The blue curve and arrows mark the shifting reflections (1,\,0,\,0)\ +\ $k_{2}$ and  (-2,\,0,\,1)\ +\ $k_{2}$, respectively. The peak positions in the intermediate region are fitted with Gaussian functions (solid lines).\ A constant vertical shift has been applied to patterns measured above $T^{*}$.\ The remnant peak above $T^{*}$ is indexed as (0.5,\,-1,\,0).\ (b) Temperature dependence of the (1.5,\,0,\,0) and (-0.5,\,0,\,1) reflections generated by $k_{1}$, which in contrast do not shift.\ (c) Temperature dependences of the \textit{x} and \textit{z} components of $k_{2}$ around $T^{*}$. C = commensurate, IC = incommensurate, and PM = paramagnetic.}
	\label{fig:2}
\end{figure}

\begin{figure}
	\centering
	\includegraphics[width=0.45\textwidth]{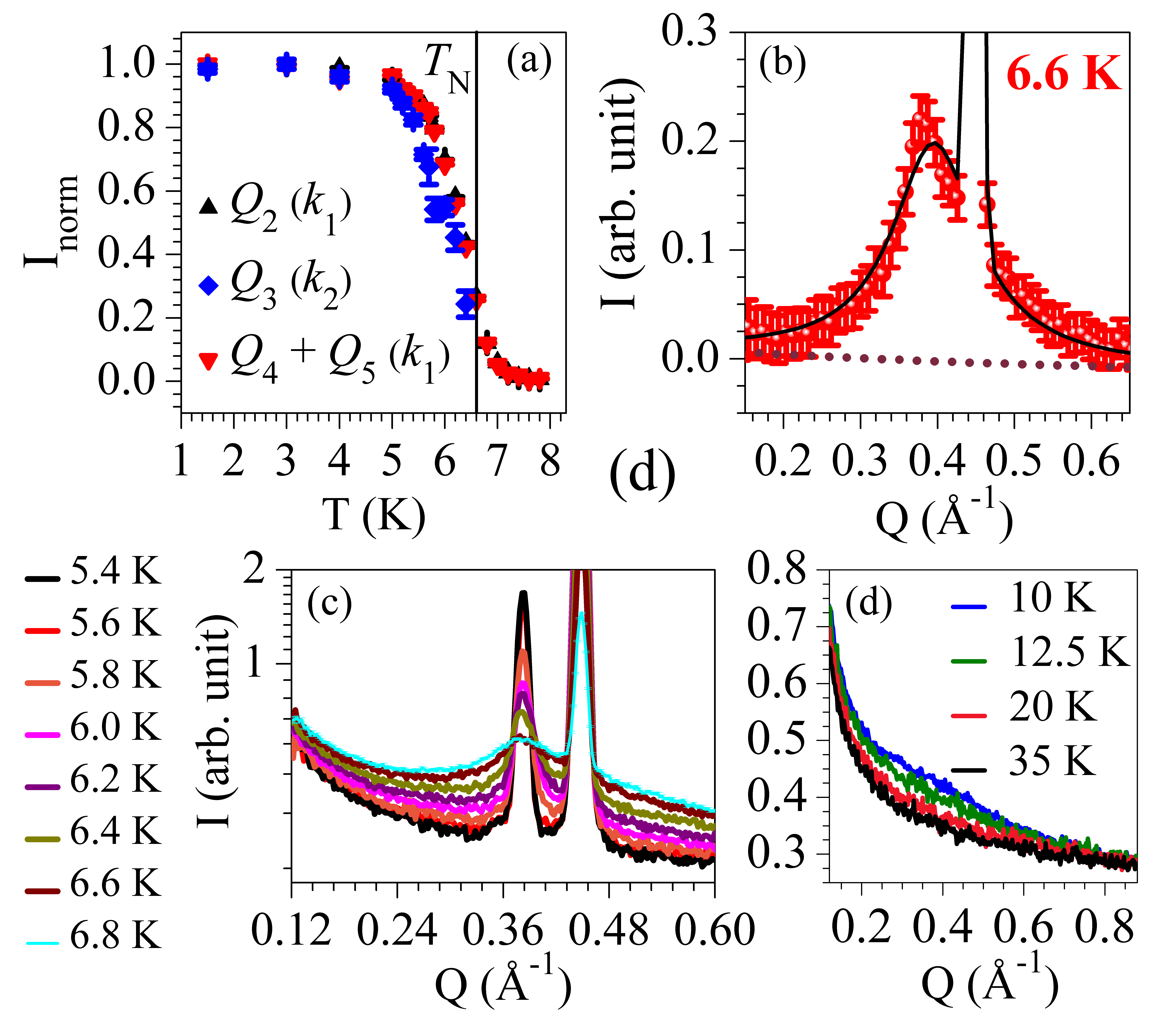}
	\caption{(a) Normalized intensity versus temperature plots of reflections at $Q_{2}$\,=\,(0.5,\,0,\,0) and $Q_{3}$\,=\,(1,\,0,\,0)\,+\,$k_{2}$, and the magnetic Bragg peak $\sim$\,1.35\,\AA$^{-1}$ consisting of $Q_{4}$\,=\,(1.5,\,0,\,0) and $Q_{5}$\,=\,(-0.5,\,0,\,1) reflections.\ (b) Intensity versus \textit{Q} curve around the $Q_{1}$\,=\,(-0.25,\,0,\,0.25) and the $Q_{2}$ reflections at 6.6\,K. Nuclear scattering background,\,taken at 35\,K,\,has been subtracted.\ The solid lines are numerical fits described in the text.\ (c)\,-\,(d) Evolution of the diffuse scattering signals in the low-\textit{Q} region as a function of temperature.}
	\label{fig:3}
\end{figure}

We have also investigated the temperature dependences of the two phases.\ The magnetic reflections generated by $k_{2}$\,=\,(-0.25,\,0,\,0.25) are greatly suppressed on heating from 5.6 K (=\,$T^{*}$) to 5.8\,K. For example, the $Q_{3}$\,=\,(0.75,\,0,\,0.25) reflection at $\sim$\,0.77\,\AA$^{-1}$ can barely be resolved above \textit{T}$^{*}$, and the remnant intensity is mainly composed of the (0.5,\,-1,\,0) reflection arising from the $k_{1}$ phase\,(Fig.\,\ref{fig:1}b).\ Concomitantly, emergent reflections which cannot be indexed using either $k_{1}$\,=\,(0.5,\,0,\,0) or $k_{2}$\,=\,(-0.25,\,0,\,0.25) appear in a broad $Q$-range\,(Fig.\,\ref{fig:2}a).\ As the temperature increases further beyond \textit{T}$^{*}$, the emergent reflection on the left of (0.75,\,0,\,0.25) continuously shifts towards the low-\textit{Q} region until it falls under the strong diffuse scattering background at 6.6\,K\,(Fig.\,\ref{fig:2}a).\ By fitting 5 clearly observable emergent reflections, we can rule out the possibility of a commensurate modulation above $T^{*}$ for these reflections.\ Unfortunately, an extensive search in incommensurate space produces sets of solutions that cannot be distinguished within our resolution.\ The peak between 1.33\,\AA$^{-1}$ and 1.38\,\AA$^{-1}$ consists exclusively of $Q_{4}$\,=\,(1.5,\,0,\,0) and $Q_{5}$\,=\,(-0.5,\,0,\,1) reflections of the $k_{1}$ phase.\ Although its intensity starts to drop around $T^{*}$\,(Fig.\,\ref{fig:3}a), no additional peaks are observed around it (Fig.\,\ref{fig:2}b).\,This suggests that the appearance of the incommensurate peaks above $T^{*}$ is not related to the $k_{1}$ phase. Since previous heat capacity measurements did not reveal any phase transition at $T^{*}$\,\cite{He2,Kimber,Nandi}, these features are consistent with a commensurate-incommensurate lock-in transition of the $k_{2}$ phase.\ We find that only two of the three components of the general incommensurate wavevector, $k_{2}$\,=\,($k_{x}$,\,$k_{y}$,\,$k_{z}$), can be uniquely determined at each temperature from the 5 clearly observable incommensurate peaks. Setting $k_{y}$\,=\,0, we may plot the temperature dependence of $k_{2}$\,=\,($k_{x}$,\,0,\,$k_{z}$) in Fig.\,\ref{fig:2}c. The temperature dependence of the normalized integrated intensity of the $Q_{3}$ reflection is also plotted in Fig.\,\ref{fig:3}a.\,\textit{T}$\mathrm{_{N}}$ for the $k_{1}$ phase has been determined to be 6.6\,K (the corresponding normalized intensity versus temperature plot has the steepest slope at this point). Since no reflection indexed by $k_{2}$ can be observed above \textit{T}$\mathrm{_{N}}$, we expect that both phases share the same transition temperature. 

Strong diffuse scattering profiles appear above $T^{*}$ (Fig.\,\ref{fig:3}c), and are detectable up to 25\,K\ (Fig.\,\ref{fig:3}d). When \textit{T}\,$\geqslant$\,\textit{T}$\mathrm{_{N}}$, the magnetic incoherent scattering background is stabilized, making it possible to study the pure magnetic diffuse scattering signals by subtracting the nuclear contributions taken at 35\,K. As shown in Fig.\,\ref{fig:3}b, these profiles still center around $Q_{1}$ at \textit{T}$\mathrm{_{N}}$.\,Fitting them with a Lorentzian function produces a correlation length ($\xi$) of 94(4)\,\AA{}. This is much smaller than $\xi$\,$\sim$\,230\,\AA{} at 1.5\,K by fitting the diffuse tails of $Q_{1}$ and $Q_{2}$ reflections\,(Fig.\,\ref{fig:3}b).\ Although spin fluctuations set in well above \textit{T}$\mathrm{_{N}}$ in $\gamma$CVO, it is very hard to extract their positions at high temperatures due to the extra scattering signals from small angles as well as the weak intensities. However, these spin fluctuations are more related to the $k_{2}$ modulation, as revealed by our analysis at temperatures close to \textit{T}$\mathrm{_{N}}$. Given that the $k_{1}$ phase populates the majority ($\sim$\,65\,\%) of the sample, the dominant spin fluctuations related to $k_{2}$ above \textit{T}$\mathrm{_{N}}$ are very surprising.  
 
 \section{Discussion and Summary}
Incommensurate magnetic microphases with a metastable propagation vector have been studied theoretically on a geometrically frustrated lattice with Ising anisotropy\,\cite{kamiya}.\,At very low temperatures, the magnetic structure is commensurate, while metastable incommensurate microphases exist in the intermediate region.\,It is also suggested that additional subtle coupling terms may stabilize the incommensurate state, as realized in Ca$_{3}$Co$_{2}$O$_{6}$\,\cite{Agrestini}.\ On the other hand, both single-ion anisotropy and exchange frustration are present in both $\alpha$CVO and $\gamma$CVO\,\cite{Wallington,Kim}.\ As suggested in Refs.\:\onlinecite{Fishman,Gang,Parshall,Seabra}, the system will form a collinear spin structure if the single-ion anisotropy is stronger compared with the spin exchange interactions, whereas an incommensurate noncollinear spin structure is favoured oppositely. The collinear spin arrangement of $\alpha$CVO, which possesses a very strong SOC, is consistent with this description\,\cite{Markkula,Lenertz2,Hollmann,Wallington}. For the $\gamma$CVO compound where the SOC is much weaker\,\cite{Hollmann}, we propose it is close to the collinear-noncollinear phase boundary. Although previous neutron diffraction measurements at much shorter wavelength support a single crystallographic phase in the magnetically ordered region of $\gamma$CVO\,\cite{Kimber,Lenertz}, local disorder, which is not sensitive to these scattering techniques, may cause the \textit{2}-phase separation. However, since our diffraction patterns are essentially identical with the ones obtained previously\,\cite{Kimber,Lenertz}, the magnetic phase separation is an intrinsic feature in $\gamma$CVO. Frustrated magnets are expected to be composed of many states that are close or degenerate in energy. If they are degenerate, a liquid, glass or ice configuration will develop. If they are close in energy, the ordered magnetic moment is typically suppressed compared to the theoretical value. We propose a third possibility, namely that the system selects two energetically equivalent states, partially lifting the degeneracy. As the ordered moment observed in each of these states matches well with the theoretical expectation, the other states should be further away in energy. Of course, the next question is which interactions might be involved in stabilizing such a setup. It has been suggested that the magnetoelectric coupling is responsible for the additional ferrimagnetic microphase in Ca$_{3}$Co$_{2}$O$_{6}$\,\cite{Prsa}. We note this term is also allowed for the $k_{2}$ phase of $\gamma$CVO due to the broken inversion symmetry of the Co(2) spin lattice\,\cite{Cheong,Choi,Singh}.

The complexity of magnetism in $\gamma$CVO can be further stressed by the reported observation on single crystal samples of magnetic reflections possibly indexed by $k_{3}$\,=\,(-1/3,\,0,\,1/3) at \textit{Q}\ $\sim$ 0.52\,\AA$^{-1}$ below $T^{*}$\,\cite{Drees}, which are not seen in our study (Fig.\,\ref{fig:1}b). This discrepancy may be related to the non-equilibrium spin dynamics, as observed in the Q1D frustrated magnet Ca$_{3}$Co$_{2}$O$_{6}$\,\cite{Agrestini2}. In our powder diffraction experiment, we recorded diffraction patterns every 0.5\,-\,1.0\ hour. By checking the time dependence of the diffraction pattern at each temperature up to 6\ hours, we did not find any relaxation behaviour. As a result, the spin dynamics in $\gamma$CVO, if it exists, needs to be either faster than 0.5 hour or slower than 6 hours.

In summary, we have investigated the magnetism of $\gamma$CVO using the neutron powder diffraction technique.\,We have established that its low temperature spin structure consists of two single-\textit{k} phases in a ratio about 65(1)\,:\,35(1). As the underlying crystallographic lattice is not geometrically frustrated, the magnetic phase separation in this material is more likely caused by the complex exchange network, e.g. exchange frustration. For the minority phase, a commensurate-incommensurate lock-in transition is observed at $T^{*}$. Our results clearly provide important motivation for future studies, e.g. further theoretical treatments concerning static phase separation in frustrated magnets are demanded.  \par       
\begin{acknowledgments}
We acknowledge the UK EPSRC for funding under grant number EP/J016977/1. This work is based on experiments performed at the Swiss spallation neutron source (SINQ), Paul Scherrer Institute, Villigen, Switzerland. Z. H. thanks the National Basic Research Program of China (No. 2012CB921701) and the Chinese Academy of Sciences under Grant No. KJZD-EW-M05 for financial support. M. I. thanks the Center for Elemental Research, Tokyo Institute of Technology for financial support.
\end{acknowledgments}


\begin{thebibliography}{39}%
\makeatletter
\providecommand \@ifxundefined [1]{%
 \@ifx{#1\undefined}
}%
\providecommand \@ifnum [1]{%
 \ifnum #1\expandafter \@firstoftwo
 \else \expandafter \@secondoftwo
 \fi
}%
\providecommand \@ifx [1]{%
 \ifx #1\expandafter \@firstoftwo
 \else \expandafter \@secondoftwo
 \fi
}%
\providecommand \natexlab [1]{#1}%
\providecommand \enquote  [1]{``#1''}%
\providecommand \bibnamefont  [1]{#1}%
\providecommand \bibfnamefont [1]{#1}%
\providecommand \citenamefont [1]{#1}%
\providecommand \href@noop [0]{\@secondoftwo}%
\providecommand \href [0]{\begingroup \@sanitize@url \@href}%
\providecommand \@href[1]{\@@startlink{#1}\@@href}%
\providecommand \@@href[1]{\endgroup#1\@@endlink}%
\providecommand \@sanitize@url [0]{\catcode `\\12\catcode `\$12\catcode
  `\&12\catcode `\#12\catcode `\^12\catcode `\_12\catcode `\%12\relax}%
\providecommand \@@startlink[1]{}%
\providecommand \@@endlink[0]{}%
\providecommand \url  [0]{\begingroup\@sanitize@url \@url }%
\providecommand \@url [1]{\endgroup\@href {#1}{\urlprefix }}%
\providecommand \urlprefix  [0]{URL }%
\providecommand \Eprint [0]{\href }%
\providecommand \doibase [0]{http://dx.doi.org/}%
\providecommand \selectlanguage [0]{\@gobble}%
\providecommand \bibinfo  [0]{\@secondoftwo}%
\providecommand \bibfield  [0]{\@secondoftwo}%
\providecommand \translation [1]{[#1]}%
\providecommand \BibitemOpen [0]{}%
\providecommand \bibitemStop [0]{}%
\providecommand \bibitemNoStop [0]{.\EOS\space}%
\providecommand \EOS [0]{\spacefactor3000\relax}%
\providecommand \BibitemShut  [1]{\csname bibitem#1\endcsname}%
\let\auto@bib@innerbib\@empty
\bibitem [{\citenamefont {Seabra}\ \emph {et~al.}(2016)\citenamefont {Seabra},
  \citenamefont {Sindzingre}, \citenamefont {Momoi},\ and\ \citenamefont
  {Shannon}}]{Seabra}%
  \BibitemOpen
  \bibfield  {author} {\bibinfo {author} {\bibfnamefont {L.}~\bibnamefont
  {Seabra}}, \bibinfo {author} {\bibfnamefont {P.}~\bibnamefont {Sindzingre}},
  \bibinfo {author} {\bibfnamefont {T.}~\bibnamefont {Momoi}}, \ and\ \bibinfo
  {author} {\bibfnamefont {N.}~\bibnamefont {Shannon}},\ }\href {\doibase
  10.1103/PhysRevB.93.085132} {\bibfield  {journal} {\bibinfo  {journal} {Phys.
  Rev. B}\ }\textbf {\bibinfo {volume} {93}},\ \bibinfo {pages} {085132}
  (\bibinfo {year} {2016})}\BibitemShut {NoStop}%
\bibitem [{\citenamefont {Gardner}\ \emph {et~al.}(2010)\citenamefont
  {Gardner}, \citenamefont {Gingras},\ and\ \citenamefont {Greedan}}]{Gardner}%
  \BibitemOpen
  \bibfield  {author} {\bibinfo {author} {\bibfnamefont {J.~S.}\ \bibnamefont
  {Gardner}}, \bibinfo {author} {\bibfnamefont {M.~J.~P.}\ \bibnamefont
  {Gingras}}, \ and\ \bibinfo {author} {\bibfnamefont {J.~E.}\ \bibnamefont
  {Greedan}},\ }\href {\doibase 10.1103/RevModPhys.82.53} {\bibfield  {journal}
  {\bibinfo  {journal} {Rev. Mod. Phys.}\ }\textbf {\bibinfo {volume} {82}},\
  \bibinfo {pages} {53} (\bibinfo {year} {2010})}\BibitemShut {NoStop}%
\bibitem [{\citenamefont {Diep}(2013)}]{Diep}%
  \BibitemOpen
  \bibfield  {author} {\bibinfo {author} {\bibfnamefont {H.~T.}\ \bibnamefont
  {Diep}},\ }\href@noop {} {\emph {\bibinfo {title} {Frustrated Spin
  Systems}}},\ \bibinfo {edition} {\textit{2nd}}\ ed.\ (\bibinfo  {publisher}
  {World Scientific},\ \bibinfo {year} {2013})\BibitemShut {NoStop}%
\bibitem [{\citenamefont {Savary}\ \emph {et~al.}(2012)\citenamefont {Savary},
  \citenamefont {Ross}, \citenamefont {Gaulin}, \citenamefont {Ruff},\ and\
  \citenamefont {Balents}}]{Savary}%
  \BibitemOpen
  \bibfield  {author} {\bibinfo {author} {\bibfnamefont {L.}~\bibnamefont
  {Savary}}, \bibinfo {author} {\bibfnamefont {K.~A.}\ \bibnamefont {Ross}},
  \bibinfo {author} {\bibfnamefont {B.~D.}\ \bibnamefont {Gaulin}}, \bibinfo
  {author} {\bibfnamefont {J.~P.~C.}\ \bibnamefont {Ruff}}, \ and\ \bibinfo
  {author} {\bibfnamefont {L.}~\bibnamefont {Balents}},\ }\href {\doibase
  10.1103/PhysRevLett.109.167201} {\bibfield  {journal} {\bibinfo  {journal}
  {Phys. Rev. Lett.}\ }\textbf {\bibinfo {volume} {109}},\ \bibinfo {pages}
  {167201} (\bibinfo {year} {2012})}\BibitemShut {NoStop}%
\bibitem [{\citenamefont {Okubo}\ \emph {et~al.}(2011)\citenamefont {Okubo},
  \citenamefont {Nguyen},\ and\ \citenamefont {Kawamura}}]{Okubo1}%
  \BibitemOpen
  \bibfield  {author} {\bibinfo {author} {\bibfnamefont {T.}~\bibnamefont
  {Okubo}}, \bibinfo {author} {\bibfnamefont {T.~H.}\ \bibnamefont {Nguyen}}, \
  and\ \bibinfo {author} {\bibfnamefont {H.}~\bibnamefont {Kawamura}},\ }\href
  {\doibase 10.1103/PhysRevB.84.144432} {\bibfield  {journal} {\bibinfo
  {journal} {Phys. Rev. B}\ }\textbf {\bibinfo {volume} {84}},\ \bibinfo
  {pages} {144432} (\bibinfo {year} {2011})}\BibitemShut {NoStop}%
\bibitem [{\citenamefont {Okubo}\ \emph {et~al.}(2012)\citenamefont {Okubo},
  \citenamefont {Chung},\ and\ \citenamefont {Kawamura}}]{Okubo2}%
  \BibitemOpen
  \bibfield  {author} {\bibinfo {author} {\bibfnamefont {T.}~\bibnamefont
  {Okubo}}, \bibinfo {author} {\bibfnamefont {S.}~\bibnamefont {Chung}}, \ and\
  \bibinfo {author} {\bibfnamefont {H.}~\bibnamefont {Kawamura}},\ }\href
  {\doibase 10.1103/PhysRevLett.108.017206} {\bibfield  {journal} {\bibinfo
  {journal} {Phys. Rev. Lett.}\ }\textbf {\bibinfo {volume} {108}},\ \bibinfo
  {pages} {017206} (\bibinfo {year} {2012})}\BibitemShut {NoStop}%
\bibitem [{\citenamefont {Lee}\ \emph {et~al.}(2010)\citenamefont {Lee},
  \citenamefont {Kaul},\ and\ \citenamefont {Balents}}]{Lee}%
  \BibitemOpen
  \bibfield  {author} {\bibinfo {author} {\bibfnamefont {S.}~\bibnamefont
  {Lee}}, \bibinfo {author} {\bibfnamefont {R.~K.}\ \bibnamefont {Kaul}}, \
  and\ \bibinfo {author} {\bibfnamefont {L.}~\bibnamefont {Balents}},\
  }\href@noop {} {\bibfield  {journal} {\bibinfo  {journal} {Nat Phys}\
  }\textbf {\bibinfo {volume} {6}},\ \bibinfo {pages} {702} (\bibinfo {year}
  {2010})}\BibitemShut {NoStop}%
\bibitem [{\citenamefont {Coldea}\ \emph {et~al.}(2010)\citenamefont {Coldea},
  \citenamefont {Tennant}, \citenamefont {Wheeler}, \citenamefont {Wawrzynska},
  \citenamefont {Prabhakaran}, \citenamefont {Telling}, \citenamefont
  {Habicht}, \citenamefont {Smeibidl},\ and\ \citenamefont {Kiefer}}]{Coldea}%
  \BibitemOpen
  \bibfield  {author} {\bibinfo {author} {\bibfnamefont {R.}~\bibnamefont
  {Coldea}}, \bibinfo {author} {\bibfnamefont {D.~A.}\ \bibnamefont {Tennant}},
  \bibinfo {author} {\bibfnamefont {E.~M.}\ \bibnamefont {Wheeler}}, \bibinfo
  {author} {\bibfnamefont {E.}~\bibnamefont {Wawrzynska}}, \bibinfo {author}
  {\bibfnamefont {D.}~\bibnamefont {Prabhakaran}}, \bibinfo {author}
  {\bibfnamefont {M.}~\bibnamefont {Telling}}, \bibinfo {author} {\bibfnamefont
  {K.}~\bibnamefont {Habicht}}, \bibinfo {author} {\bibfnamefont
  {P.}~\bibnamefont {Smeibidl}}, \ and\ \bibinfo {author} {\bibfnamefont
  {K.}~\bibnamefont {Kiefer}},\ }\href {\doibase 10.1126/science.1180085}
  {\bibfield  {journal} {\bibinfo  {journal} {Science}\ }\textbf {\bibinfo
  {volume} {327}},\ \bibinfo {pages} {177} (\bibinfo {year}
  {2010})}\BibitemShut {NoStop}%
\bibitem [{\citenamefont {Dagotto}\ \emph {et~al.}(2001)\citenamefont
  {Dagotto}, \citenamefont {Hotta},\ and\ \citenamefont {Moreo}}]{Dagotto}%
  \BibitemOpen
  \bibfield  {author} {\bibinfo {author} {\bibfnamefont {E.}~\bibnamefont
  {Dagotto}}, \bibinfo {author} {\bibfnamefont {T.}~\bibnamefont {Hotta}}, \
  and\ \bibinfo {author} {\bibfnamefont {A.}~\bibnamefont {Moreo}},\ }\href
  {\doibase http://dx.doi.org/10.1016/S0370-1573(00)00121-6} {\bibfield
  {journal} {\bibinfo  {journal} {Physics Reports}\ }\textbf {\bibinfo {volume}
  {344}},\ \bibinfo {pages} {1 } (\bibinfo {year} {2001})}\BibitemShut
  {NoStop}%
\bibitem [{\citenamefont {Emery}\ and\ \citenamefont {Kivelson}(1993)}]{EMERY}%
  \BibitemOpen
  \bibfield  {author} {\bibinfo {author} {\bibfnamefont {V.}~\bibnamefont
  {Emery}}\ and\ \bibinfo {author} {\bibfnamefont {S.}~\bibnamefont
  {Kivelson}},\ }\href {\doibase
  http://dx.doi.org/10.1016/0921-4534(93)90581-A} {\bibfield  {journal}
  {\bibinfo  {journal} {Physica C: Superconductivity}\ }\textbf {\bibinfo
  {volume} {209}},\ \bibinfo {pages} {597 } (\bibinfo {year}
  {1993})}\BibitemShut {NoStop}%
\bibitem [{\citenamefont {Matsuda}\ \emph {et~al.}(2015)\citenamefont
  {Matsuda}, \citenamefont {Partzsch}, \citenamefont {Tsuyama}, \citenamefont
  {Schierle}, \citenamefont {Weschke}, \citenamefont {Geck}, \citenamefont
  {Saito}, \citenamefont {Ishiwata}, \citenamefont {Tokura},\ and\
  \citenamefont {Wadati}}]{Matsuda}%
  \BibitemOpen
  \bibfield  {author} {\bibinfo {author} {\bibfnamefont {T.}~\bibnamefont
  {Matsuda}}, \bibinfo {author} {\bibfnamefont {S.}~\bibnamefont {Partzsch}},
  \bibinfo {author} {\bibfnamefont {T.}~\bibnamefont {Tsuyama}}, \bibinfo
  {author} {\bibfnamefont {E.}~\bibnamefont {Schierle}}, \bibinfo {author}
  {\bibfnamefont {E.}~\bibnamefont {Weschke}}, \bibinfo {author} {\bibfnamefont
  {J.}~\bibnamefont {Geck}}, \bibinfo {author} {\bibfnamefont {T.}~\bibnamefont
  {Saito}}, \bibinfo {author} {\bibfnamefont {S.}~\bibnamefont {Ishiwata}},
  \bibinfo {author} {\bibfnamefont {Y.}~\bibnamefont {Tokura}}, \ and\ \bibinfo
  {author} {\bibfnamefont {H.}~\bibnamefont {Wadati}},\ }\href {\doibase
  10.1103/PhysRevLett.114.236403} {\bibfield  {journal} {\bibinfo  {journal}
  {Phys. Rev. Lett.}\ }\textbf {\bibinfo {volume} {114}},\ \bibinfo {pages}
  {236403} (\bibinfo {year} {2015})}\BibitemShut {NoStop}%
\bibitem [{\citenamefont {Agrestini}\ \emph {et~al.}(2011)\citenamefont
  {Agrestini}, \citenamefont {Fleck}, \citenamefont {Chapon}, \citenamefont
  {Mazzoli}, \citenamefont {Bombardi}, \citenamefont {Lees},\ and\
  \citenamefont {Petrenko}}]{Agrestini2}%
  \BibitemOpen
  \bibfield  {author} {\bibinfo {author} {\bibfnamefont {S.}~\bibnamefont
  {Agrestini}}, \bibinfo {author} {\bibfnamefont {C.~L.}\ \bibnamefont
  {Fleck}}, \bibinfo {author} {\bibfnamefont {L.~C.}\ \bibnamefont {Chapon}},
  \bibinfo {author} {\bibfnamefont {C.}~\bibnamefont {Mazzoli}}, \bibinfo
  {author} {\bibfnamefont {A.}~\bibnamefont {Bombardi}}, \bibinfo {author}
  {\bibfnamefont {M.~R.}\ \bibnamefont {Lees}}, \ and\ \bibinfo {author}
  {\bibfnamefont {O.~A.}\ \bibnamefont {Petrenko}},\ }\href {\doibase
  10.1103/PhysRevLett.106.197204} {\bibfield  {journal} {\bibinfo  {journal}
  {Phys. Rev. Lett.}\ }\textbf {\bibinfo {volume} {106}},\ \bibinfo {pages}
  {197204} (\bibinfo {year} {2011})}\BibitemShut {NoStop}%
\bibitem [{\citenamefont {Kamiya}\ and\ \citenamefont
  {Batista}(2012)}]{kamiya}%
  \BibitemOpen
  \bibfield  {author} {\bibinfo {author} {\bibfnamefont {Y.}~\bibnamefont
  {Kamiya}}\ and\ \bibinfo {author} {\bibfnamefont {C.~D.}\ \bibnamefont
  {Batista}},\ }\href {\doibase 10.1103/PhysRevLett.109.067204} {\bibfield
  {journal} {\bibinfo  {journal} {Phys. Rev. Lett.}\ }\textbf {\bibinfo
  {volume} {109}},\ \bibinfo {pages} {067204} (\bibinfo {year}
  {2012})}\BibitemShut {NoStop}%
\bibitem [{\citenamefont {Prsa}\ \emph {et~al.}(2014)\citenamefont {Prsa},
  \citenamefont {Laver}, \citenamefont {Mansson},\ and\ \citenamefont
  {et~al}}]{Prsa}%
  \BibitemOpen
  \bibfield  {author} {\bibinfo {author} {\bibfnamefont {K.}~\bibnamefont
  {Prsa}}, \bibinfo {author} {\bibfnamefont {M.}~\bibnamefont {Laver}},
  \bibinfo {author} {\bibfnamefont {M.}~\bibnamefont {Mansson}}, \ and\
  \bibinfo {author} {\bibnamefont {et~al}},\ }\href@noop {} {\enquote {\bibinfo
  {title} {Magnetic nano-fluctuations in a frustrated magnet},}\ } (\bibinfo
  {year} {2014}),\ \Eprint {http://arxiv.org/abs/1404.7398} {arXiv:1404.7398}
  \BibitemShut {NoStop}%
\bibitem [{\citenamefont {Momma}\ and\ \citenamefont {Izumi}(2011)}]{VESTA}%
  \BibitemOpen
  \bibfield  {author} {\bibinfo {author} {\bibfnamefont {K.}~\bibnamefont
  {Momma}}\ and\ \bibinfo {author} {\bibfnamefont {F.}~\bibnamefont {Izumi}},\
  }\href {\doibase 10.1107/S0021889811038970} {\bibfield  {journal} {\bibinfo
  {journal} {Journal of Applied Crystallography}\ }\textbf {\bibinfo {volume}
  {44}},\ \bibinfo {pages} {1272} (\bibinfo {year} {2011})}\BibitemShut
  {NoStop}%
\bibitem [{\citenamefont {MÃŒller-Buschbaum}\ and\ \citenamefont
  {Kobel}(1991)}]{MULLER}%
  \BibitemOpen
  \bibfield  {author} {\bibinfo {author} {\bibfnamefont {H.}~\bibnamefont
  {MÃŒller-Buschbaum}}\ and\ \bibinfo {author} {\bibfnamefont {M.}~\bibnamefont
  {Kobel}},\ }\href {\doibase http://dx.doi.org/10.1016/0925-8388(91)90008-J}
  {\bibfield  {journal} {\bibinfo  {journal} {Journal of Alloys and Compounds}\
  }\textbf {\bibinfo {volume} {176}},\ \bibinfo {pages} {39 } (\bibinfo {year}
  {1991})}\BibitemShut {NoStop}%
\bibitem [{\citenamefont {He}\ \emph {et~al.}(2009)\citenamefont {He},
  \citenamefont {Yamaura}, \citenamefont {Ueda},\ and\ \citenamefont
  {Cheng}}]{He}%
  \BibitemOpen
  \bibfield  {author} {\bibinfo {author} {\bibfnamefont {Z.}~\bibnamefont
  {He}}, \bibinfo {author} {\bibfnamefont {J.-I.}\ \bibnamefont {Yamaura}},
  \bibinfo {author} {\bibfnamefont {Y.}~\bibnamefont {Ueda}}, \ and\ \bibinfo
  {author} {\bibfnamefont {W.}~\bibnamefont {Cheng}},\ }\href {\doibase
  10.1021/ja902623b} {\bibfield  {journal} {\bibinfo  {journal} {Journal of the
  American Chemical Society}\ }\textbf {\bibinfo {volume} {131}},\ \bibinfo
  {pages} {7554} (\bibinfo {year} {2009})}\BibitemShut {NoStop}%
\bibitem [{\citenamefont {He}\ and\ \citenamefont {Itoh}(2014)}]{He2}%
  \BibitemOpen
  \bibfield  {author} {\bibinfo {author} {\bibfnamefont {Z.}~\bibnamefont
  {He}}\ and\ \bibinfo {author} {\bibfnamefont {M.}~\bibnamefont {Itoh}},\
  }\href {\doibase http://dx.doi.org/10.1016/j.jcrysgro.2013.11.078} {\bibfield
   {journal} {\bibinfo  {journal} {Journal of Crystal Growth}\ }\textbf
  {\bibinfo {volume} {388}},\ \bibinfo {pages} {103 } (\bibinfo {year}
  {2014})}\BibitemShut {NoStop}%
\bibitem [{\citenamefont {Kimber}\ \emph {et~al.}(2011)\citenamefont {Kimber},
  \citenamefont {Mutka}, \citenamefont {Chatterji}, \citenamefont {Hofmann},
  \citenamefont {Henry}, \citenamefont {Bordallo}, \citenamefont {Argyriou},\
  and\ \citenamefont {Attfield}}]{Kimber}%
  \BibitemOpen
  \bibfield  {author} {\bibinfo {author} {\bibfnamefont {S.~A.~J.}\
  \bibnamefont {Kimber}}, \bibinfo {author} {\bibfnamefont {H.}~\bibnamefont
  {Mutka}}, \bibinfo {author} {\bibfnamefont {T.}~\bibnamefont {Chatterji}},
  \bibinfo {author} {\bibfnamefont {T.}~\bibnamefont {Hofmann}}, \bibinfo
  {author} {\bibfnamefont {P.~F.}\ \bibnamefont {Henry}}, \bibinfo {author}
  {\bibfnamefont {H.~N.}\ \bibnamefont {Bordallo}}, \bibinfo {author}
  {\bibfnamefont {D.~N.}\ \bibnamefont {Argyriou}}, \ and\ \bibinfo {author}
  {\bibfnamefont {J.~P.}\ \bibnamefont {Attfield}},\ }\href {\doibase
  10.1103/PhysRevB.84.104425} {\bibfield  {journal} {\bibinfo  {journal} {Phys.
  Rev. B}\ }\textbf {\bibinfo {volume} {84}},\ \bibinfo {pages} {104425}
  (\bibinfo {year} {2011})}\BibitemShut {NoStop}%
\bibitem [{Ful()}]{FullProf}%
  \BibitemOpen
  \href@noop {} {\ }\bibinfo {note} {FullProf Suite,
  https://www.ill.eu/sites/fullprof}\BibitemShut {NoStop}%
\bibitem [{\citenamefont {Lenertz}\ \emph {et~al.}(2014)\citenamefont
  {Lenertz}, \citenamefont {Dinia}, \citenamefont {Colis}, \citenamefont
  {MentrÃ©}, \citenamefont {AndrÃ©}, \citenamefont {Porcher},\ and\
  \citenamefont {Suard}}]{Lenertz}%
  \BibitemOpen
  \bibfield  {author} {\bibinfo {author} {\bibfnamefont {M.}~\bibnamefont
  {Lenertz}}, \bibinfo {author} {\bibfnamefont {A.}~\bibnamefont {Dinia}},
  \bibinfo {author} {\bibfnamefont {S.}~\bibnamefont {Colis}}, \bibinfo
  {author} {\bibfnamefont {O.}~\bibnamefont {MentrÃ©}}, \bibinfo {author}
  {\bibfnamefont {G.}~\bibnamefont {AndrÃ©}}, \bibinfo {author} {\bibfnamefont
  {F.}~\bibnamefont {Porcher}}, \ and\ \bibinfo {author} {\bibfnamefont
  {E.}~\bibnamefont {Suard}},\ }\href {\doibase 10.1021/jp503389c} {\bibfield
  {journal} {\bibinfo  {journal} {The Journal of Physical Chemistry C}\
  }\textbf {\bibinfo {volume} {118}},\ \bibinfo {pages} {13981} (\bibinfo
  {year} {2014})}\BibitemShut {NoStop}%
\bibitem [{\citenamefont {Stewart}\ \emph {et~al.}(2004)\citenamefont
  {Stewart}, \citenamefont {Ehlers}, \citenamefont {Wills}, \citenamefont
  {Bramwell},\ and\ \citenamefont {Gardner}}]{Stewart}%
  \BibitemOpen
  \bibfield  {author} {\bibinfo {author} {\bibfnamefont {J.~R.}\ \bibnamefont
  {Stewart}}, \bibinfo {author} {\bibfnamefont {G.}~\bibnamefont {Ehlers}},
  \bibinfo {author} {\bibfnamefont {A.~S.}\ \bibnamefont {Wills}}, \bibinfo
  {author} {\bibfnamefont {S.~T.}\ \bibnamefont {Bramwell}}, \ and\ \bibinfo
  {author} {\bibfnamefont {J.~S.}\ \bibnamefont {Gardner}},\ }\href
  {http://stacks.iop.org/0953-8984/16/i=28/a=L01} {\bibfield  {journal}
  {\bibinfo  {journal} {Journal of Physics: Condensed Matter}\ }\textbf
  {\bibinfo {volume} {16}},\ \bibinfo {pages} {L321} (\bibinfo {year}
  {2004})}\BibitemShut {NoStop}%
\bibitem [{\citenamefont {Javanparast}\ \emph {et~al.}(2015)\citenamefont
  {Javanparast}, \citenamefont {Hao}, \citenamefont {Enjalran},\ and\
  \citenamefont {Gingras}}]{Javanparast}%
  \BibitemOpen
  \bibfield  {author} {\bibinfo {author} {\bibfnamefont {B.}~\bibnamefont
  {Javanparast}}, \bibinfo {author} {\bibfnamefont {Z.}~\bibnamefont {Hao}},
  \bibinfo {author} {\bibfnamefont {M.}~\bibnamefont {Enjalran}}, \ and\
  \bibinfo {author} {\bibfnamefont {M.~J.~P.}\ \bibnamefont {Gingras}},\ }\href
  {\doibase 10.1103/PhysRevLett.114.130601} {\bibfield  {journal} {\bibinfo
  {journal} {Phys. Rev. Lett.}\ }\textbf {\bibinfo {volume} {114}},\ \bibinfo
  {pages} {130601} (\bibinfo {year} {2015})}\BibitemShut {NoStop}%
\bibitem [{\citenamefont {Bourgeois}\ \emph {et~al.}(2012)\citenamefont
  {Bourgeois}, \citenamefont {Andr\'e}, \citenamefont {Petit}, \citenamefont
  {Robert}, \citenamefont {Poienar}, \citenamefont {Rouquette}, \citenamefont
  {Elka\"{\i}m}, \citenamefont {Hervieu}, \citenamefont {Maignan},
  \citenamefont {Martin},\ and\ \citenamefont {Damay}}]{Bourgeois}%
  \BibitemOpen
  \bibfield  {author} {\bibinfo {author} {\bibfnamefont {J.}~\bibnamefont
  {Bourgeois}}, \bibinfo {author} {\bibfnamefont {G.}~\bibnamefont {Andr\'e}},
  \bibinfo {author} {\bibfnamefont {S.}~\bibnamefont {Petit}}, \bibinfo
  {author} {\bibfnamefont {J.}~\bibnamefont {Robert}}, \bibinfo {author}
  {\bibfnamefont {M.}~\bibnamefont {Poienar}}, \bibinfo {author} {\bibfnamefont
  {J.}~\bibnamefont {Rouquette}}, \bibinfo {author} {\bibfnamefont
  {E.}~\bibnamefont {Elka\"{\i}m}}, \bibinfo {author} {\bibfnamefont
  {M.}~\bibnamefont {Hervieu}}, \bibinfo {author} {\bibfnamefont
  {A.}~\bibnamefont {Maignan}}, \bibinfo {author} {\bibfnamefont
  {C.}~\bibnamefont {Martin}}, \ and\ \bibinfo {author} {\bibfnamefont
  {F.}~\bibnamefont {Damay}},\ }\href {\doibase 10.1103/PhysRevB.86.024413}
  {\bibfield  {journal} {\bibinfo  {journal} {Phys. Rev. B}\ }\textbf {\bibinfo
  {volume} {86}},\ \bibinfo {pages} {024413} (\bibinfo {year}
  {2012})}\BibitemShut {NoStop}%
\bibitem [{\citenamefont {Pregelj}\ \emph {et~al.}(2015)\citenamefont
  {Pregelj}, \citenamefont {Zorko}, \citenamefont {Zaharko}, \citenamefont
  {Nojiri}, \citenamefont {Berger}, \citenamefont {Chapon},\ and\ \citenamefont
  {ArÄon}}]{Pregelj}%
  \BibitemOpen
  \bibfield  {author} {\bibinfo {author} {\bibfnamefont {M.}~\bibnamefont
  {Pregelj}}, \bibinfo {author} {\bibfnamefont {A.}~\bibnamefont {Zorko}},
  \bibinfo {author} {\bibfnamefont {O.}~\bibnamefont {Zaharko}}, \bibinfo
  {author} {\bibfnamefont {H.}~\bibnamefont {Nojiri}}, \bibinfo {author}
  {\bibfnamefont {H.}~\bibnamefont {Berger}}, \bibinfo {author} {\bibfnamefont
  {L.~C.}\ \bibnamefont {Chapon}}, \ and\ \bibinfo {author} {\bibfnamefont
  {D.}~\bibnamefont {Ar\^con}},\ }\href {\doibase 10.1038/ncomms8255} {\bibfield
   {journal} {\bibinfo  {journal} {Nat. Commun.}\ }\textbf {\bibinfo {volume}
  {6}},\ \bibinfo {pages} {7255} (\bibinfo {year} {2015})}\BibitemShut
  {NoStop}%
\bibitem [{\citenamefont {Drees}\ \emph {et~al.}(2015)\citenamefont {Drees},
  \citenamefont {Agrestini}, \citenamefont {Zaharko},\ and\ \citenamefont
  {Komarek}}]{Drees}%
  \BibitemOpen
  \bibfield  {author} {\bibinfo {author} {\bibfnamefont {Y.}~\bibnamefont
  {Drees}}, \bibinfo {author} {\bibfnamefont {S.}~\bibnamefont {Agrestini}},
  \bibinfo {author} {\bibfnamefont {O.}~\bibnamefont {Zaharko}}, \ and\
  \bibinfo {author} {\bibfnamefont {A.~C.}\ \bibnamefont {Komarek}},\ }\href
  {\doibase 10.1021/cg5015303} {\bibfield  {journal} {\bibinfo  {journal}
  {Crystal Growth \& Design}\ }\textbf {\bibinfo {volume} {15}},\ \bibinfo
  {pages} {1168} (\bibinfo {year} {2015})}\BibitemShut {NoStop}%
\bibitem [{\citenamefont {Lenertz}\ \emph {et~al.}(2012)\citenamefont
  {Lenertz}, \citenamefont {Alaria}, \citenamefont {Stoeffler}, \citenamefont
  {Colis}, \citenamefont {Dinia}, \citenamefont {Mentr\'e}, \citenamefont
  {Andr\'e}, \citenamefont {Porcher},\ and\ \citenamefont {Suard}}]{Lenertz2}%
  \BibitemOpen
  \bibfield  {author} {\bibinfo {author} {\bibfnamefont {M.}~\bibnamefont
  {Lenertz}}, \bibinfo {author} {\bibfnamefont {J.}~\bibnamefont {Alaria}},
  \bibinfo {author} {\bibfnamefont {D.}~\bibnamefont {Stoeffler}}, \bibinfo
  {author} {\bibfnamefont {S.}~\bibnamefont {Colis}}, \bibinfo {author}
  {\bibfnamefont {A.}~\bibnamefont {Dinia}}, \bibinfo {author} {\bibfnamefont
  {O.}~\bibnamefont {Mentr\'e}}, \bibinfo {author} {\bibfnamefont
  {G.}~\bibnamefont {Andr\'e}}, \bibinfo {author} {\bibfnamefont
  {F.}~\bibnamefont {Porcher}}, \ and\ \bibinfo {author} {\bibfnamefont
  {E.}~\bibnamefont {Suard}},\ }\href {\doibase 10.1103/PhysRevB.86.214428}
  {\bibfield  {journal} {\bibinfo  {journal} {Phys. Rev. B}\ }\textbf {\bibinfo
  {volume} {86}},\ \bibinfo {pages} {214428} (\bibinfo {year}
  {2012})}\BibitemShut {NoStop}%
\bibitem [{\citenamefont {Markkula}\ \emph {et~al.}(2012)\citenamefont
  {Markkula}, \citenamefont {Arevalo-Lopez},\ and\ \citenamefont
  {Attfield}}]{Markkula}%
  \BibitemOpen
  \bibfield  {author} {\bibinfo {author} {\bibfnamefont {M.}~\bibnamefont
  {Markkula}}, \bibinfo {author} {\bibfnamefont {A.~M.}\ \bibnamefont
  {Arevalo-Lopez}}, \ and\ \bibinfo {author} {\bibfnamefont {J.~P.}\
  \bibnamefont {Attfield}},\ }\href {\doibase
  http://dx.doi.org/10.1016/j.jssc.2012.04.029} {\bibfield  {journal} {\bibinfo
   {journal} {Journal of Solid State Chemistry}\ }\textbf {\bibinfo {volume}
  {192}},\ \bibinfo {pages} {390 } (\bibinfo {year} {2012})}\BibitemShut
  {NoStop}%
\bibitem [{\citenamefont {Hollmann}\ \emph {et~al.}(2014)\citenamefont
  {Hollmann}, \citenamefont {Agrestini}, \citenamefont {Hu}, \citenamefont
  {He}, \citenamefont {Schmidt}, \citenamefont {Kuo}, \citenamefont {Rotter},
  \citenamefont {Nugroho}, \citenamefont {Sessi}, \citenamefont {Tanaka},
  \citenamefont {Brookes},\ and\ \citenamefont {Tjeng}}]{Hollmann}%
  \BibitemOpen
  \bibfield  {author} {\bibinfo {author} {\bibfnamefont {N.}~\bibnamefont
  {Hollmann}}, \bibinfo {author} {\bibfnamefont {S.}~\bibnamefont {Agrestini}},
  \bibinfo {author} {\bibfnamefont {Z.}~\bibnamefont {Hu}}, \bibinfo {author}
  {\bibfnamefont {Z.}~\bibnamefont {He}}, \bibinfo {author} {\bibfnamefont
  {M.}~\bibnamefont {Schmidt}}, \bibinfo {author} {\bibfnamefont {C.-Y.}\
  \bibnamefont {Kuo}}, \bibinfo {author} {\bibfnamefont {M.}~\bibnamefont
  {Rotter}}, \bibinfo {author} {\bibfnamefont {A.~A.}\ \bibnamefont {Nugroho}},
  \bibinfo {author} {\bibfnamefont {V.}~\bibnamefont {Sessi}}, \bibinfo
  {author} {\bibfnamefont {A.}~\bibnamefont {Tanaka}}, \bibinfo {author}
  {\bibfnamefont {N.~B.}\ \bibnamefont {Brookes}}, \ and\ \bibinfo {author}
  {\bibfnamefont {L.~H.}\ \bibnamefont {Tjeng}},\ }\href {\doibase
  10.1103/PhysRevB.89.201101} {\bibfield  {journal} {\bibinfo  {journal} {Phys.
  Rev. B}\ }\textbf {\bibinfo {volume} {89}},\ \bibinfo {pages} {201101}
  (\bibinfo {year} {2014})}\BibitemShut {NoStop}%
\bibitem [{\citenamefont {Wallington}\ \emph {et~al.}(2015)\citenamefont
  {Wallington}, \citenamefont {Arevalo-Lopez}, \citenamefont {Taylor},
  \citenamefont {Stewart}, \citenamefont {Garcia-Sakai}, \citenamefont
  {Attfield},\ and\ \citenamefont {Stock}}]{Wallington}%
  \BibitemOpen
  \bibfield  {author} {\bibinfo {author} {\bibfnamefont {F.}~\bibnamefont
  {Wallington}}, \bibinfo {author} {\bibfnamefont {A.~M.}\ \bibnamefont
  {Arevalo-Lopez}}, \bibinfo {author} {\bibfnamefont {J.~W.}\ \bibnamefont
  {Taylor}}, \bibinfo {author} {\bibfnamefont {J.~R.}\ \bibnamefont {Stewart}},
  \bibinfo {author} {\bibfnamefont {V.}~\bibnamefont {Garcia-Sakai}}, \bibinfo
  {author} {\bibfnamefont {J.~P.}\ \bibnamefont {Attfield}}, \ and\ \bibinfo
  {author} {\bibfnamefont {C.}~\bibnamefont {Stock}},\ }\href {\doibase
  10.1103/PhysRevB.92.125116} {\bibfield  {journal} {\bibinfo  {journal} {Phys.
  Rev. B}\ }\textbf {\bibinfo {volume} {92}},\ \bibinfo {pages} {125116}
  (\bibinfo {year} {2015})}\BibitemShut {NoStop}%
\bibitem [{\citenamefont {Kim}\ \emph {et~al.}(2012)\citenamefont {Kim},
  \citenamefont {Kim}, \citenamefont {Kim}, \citenamefont {Choi}, \citenamefont
  {Park}, \citenamefont {Jeong},\ and\ \citenamefont {Min}}]{Kim}%
  \BibitemOpen
  \bibfield  {author} {\bibinfo {author} {\bibfnamefont {B.}~\bibnamefont
  {Kim}}, \bibinfo {author} {\bibfnamefont {B.~H.}\ \bibnamefont {Kim}},
  \bibinfo {author} {\bibfnamefont {K.}~\bibnamefont {Kim}}, \bibinfo {author}
  {\bibfnamefont {H.~C.}\ \bibnamefont {Choi}}, \bibinfo {author}
  {\bibfnamefont {S.-Y.}\ \bibnamefont {Park}}, \bibinfo {author}
  {\bibfnamefont {Y.~H.}\ \bibnamefont {Jeong}}, \ and\ \bibinfo {author}
  {\bibfnamefont {B.~I.}\ \bibnamefont {Min}},\ }\href {\doibase
  10.1103/PhysRevB.85.220407} {\bibfield  {journal} {\bibinfo  {journal} {Phys.
  Rev. B}\ }\textbf {\bibinfo {volume} {85}},\ \bibinfo {pages} {220407}
  (\bibinfo {year} {2012})}\BibitemShut {NoStop}%
\bibitem [{\citenamefont {Nandi}\ and\ \citenamefont {Mandal}(2016)}]{Nandi}%
  \BibitemOpen
  \bibfield  {author} {\bibinfo {author} {\bibfnamefont {M.}~\bibnamefont
  {Nandi}}\ and\ \bibinfo {author} {\bibfnamefont {P.}~\bibnamefont {Mandal}},\
  }\href
  {http://scitation.aip.org/content/aip/journal/jap/119/13/10.1063/1.4945395}
  {\bibfield  {journal} {\bibinfo  {journal} {Journal of Applied Physics}\
  }\textbf {\bibinfo {volume} {119}},\ \bibinfo {eid} {133904} (\bibinfo {year}
  {2016})}\BibitemShut {NoStop}%
\bibitem [{\citenamefont {Agrestini}\ \emph {et~al.}(2008)\citenamefont
  {Agrestini}, \citenamefont {Chapon}, \citenamefont {Daoud-Aladine},
  \citenamefont {Schefer}, \citenamefont {Gukasov}, \citenamefont {Mazzoli},
  \citenamefont {Lees},\ and\ \citenamefont {Petrenko}}]{Agrestini}%
  \BibitemOpen
  \bibfield  {author} {\bibinfo {author} {\bibfnamefont {S.}~\bibnamefont
  {Agrestini}}, \bibinfo {author} {\bibfnamefont {L.~C.}\ \bibnamefont
  {Chapon}}, \bibinfo {author} {\bibfnamefont {A.}~\bibnamefont
  {Daoud-Aladine}}, \bibinfo {author} {\bibfnamefont {J.}~\bibnamefont
  {Schefer}}, \bibinfo {author} {\bibfnamefont {A.}~\bibnamefont {Gukasov}},
  \bibinfo {author} {\bibfnamefont {C.}~\bibnamefont {Mazzoli}}, \bibinfo
  {author} {\bibfnamefont {M.~R.}\ \bibnamefont {Lees}}, \ and\ \bibinfo
  {author} {\bibfnamefont {O.~A.}\ \bibnamefont {Petrenko}},\ }\href {\doibase
  10.1103/PhysRevLett.101.097207} {\bibfield  {journal} {\bibinfo  {journal}
  {Phys. Rev. Lett.}\ }\textbf {\bibinfo {volume} {101}},\ \bibinfo {pages}
  {097207} (\bibinfo {year} {2008})}\BibitemShut {NoStop}%
\bibitem [{\citenamefont {Fishman}(2011)}]{Fishman}%
  \BibitemOpen
  \bibfield  {author} {\bibinfo {author} {\bibfnamefont {R.~S.}\ \bibnamefont
  {Fishman}},\ }\href {\doibase 10.1103/PhysRevLett.106.037206} {\bibfield
  {journal} {\bibinfo  {journal} {Phys. Rev. Lett.}\ }\textbf {\bibinfo
  {volume} {106}},\ \bibinfo {pages} {037206} (\bibinfo {year}
  {2011})}\BibitemShut {NoStop}%
\bibitem [{\citenamefont {Chen}\ \emph {et~al.}(2013)\citenamefont {Chen},
  \citenamefont {Choi},\ and\ \citenamefont {Radzihovsky}}]{Gang}%
  \BibitemOpen
  \bibfield  {author} {\bibinfo {author} {\bibfnamefont {G.}~\bibnamefont
  {Chen}}, \bibinfo {author} {\bibfnamefont {S.}~\bibnamefont {Choi}}, \ and\
  \bibinfo {author} {\bibfnamefont {L.}~\bibnamefont {Radzihovsky}},\ }\href
  {\doibase 10.1103/PhysRevB.88.165117} {\bibfield  {journal} {\bibinfo
  {journal} {Phys. Rev. B}\ }\textbf {\bibinfo {volume} {88}},\ \bibinfo
  {pages} {165117} (\bibinfo {year} {2013})}\BibitemShut {NoStop}%
\bibitem [{\citenamefont {Parshall}\ \emph {et~al.}(2012)\citenamefont
  {Parshall}, \citenamefont {Chen}, \citenamefont {Pintschovius}, \citenamefont
  {Lamago}, \citenamefont {Wolf}, \citenamefont {Radzihovsky},\ and\
  \citenamefont {Reznik}}]{Parshall}%
  \BibitemOpen
  \bibfield  {author} {\bibinfo {author} {\bibfnamefont {D.}~\bibnamefont
  {Parshall}}, \bibinfo {author} {\bibfnamefont {G.}~\bibnamefont {Chen}},
  \bibinfo {author} {\bibfnamefont {L.}~\bibnamefont {Pintschovius}}, \bibinfo
  {author} {\bibfnamefont {D.}~\bibnamefont {Lamago}}, \bibinfo {author}
  {\bibfnamefont {T.}~\bibnamefont {Wolf}}, \bibinfo {author} {\bibfnamefont
  {L.}~\bibnamefont {Radzihovsky}}, \ and\ \bibinfo {author} {\bibfnamefont
  {D.}~\bibnamefont {Reznik}},\ }\href {\doibase 10.1103/PhysRevB.85.140515}
  {\bibfield  {journal} {\bibinfo  {journal} {Phys. Rev. B}\ }\textbf {\bibinfo
  {volume} {85}},\ \bibinfo {pages} {140515} (\bibinfo {year}
  {2012})}\BibitemShut {NoStop}%
\bibitem [{\citenamefont {Cheong}\ and\ \citenamefont
  {Mostovoy}(2007)}]{Cheong}%
  \BibitemOpen
  \bibfield  {author} {\bibinfo {author} {\bibfnamefont {S.-W.}\ \bibnamefont
  {Cheong}}\ and\ \bibinfo {author} {\bibfnamefont {M.}~\bibnamefont
  {Mostovoy}},\ }\href {\doibase 10.1038/nmat1804} {\bibfield  {journal}
  {\bibinfo  {journal} {Nat. Mat.}\ }\textbf {\bibinfo {volume} {6}},\ \bibinfo
  {pages} {13} (\bibinfo {year} {2007})}\BibitemShut {NoStop}%
\bibitem [{\citenamefont {Choi}\ \emph {et~al.}(2008)\citenamefont {Choi},
  \citenamefont {Yi}, \citenamefont {Lee}, \citenamefont {Huang}, \citenamefont
  {Kiryukhin},\ and\ \citenamefont {Cheong}}]{Choi}%
  \BibitemOpen
  \bibfield  {author} {\bibinfo {author} {\bibfnamefont {Y.~J.}\ \bibnamefont
  {Choi}}, \bibinfo {author} {\bibfnamefont {H.~T.}\ \bibnamefont {Yi}},
  \bibinfo {author} {\bibfnamefont {S.}~\bibnamefont {Lee}}, \bibinfo {author}
  {\bibfnamefont {Q.}~\bibnamefont {Huang}}, \bibinfo {author} {\bibfnamefont
  {V.}~\bibnamefont {Kiryukhin}}, \ and\ \bibinfo {author} {\bibfnamefont
  {S.-W.}\ \bibnamefont {Cheong}},\ }\href {\doibase
  10.1103/PhysRevLett.100.047601} {\bibfield  {journal} {\bibinfo  {journal}
  {Phys. Rev. Lett.}\ }\textbf {\bibinfo {volume} {100}},\ \bibinfo {pages}
  {047601} (\bibinfo {year} {2008})}\BibitemShut {NoStop}%
\bibitem [{\citenamefont {Singh}\ \emph {et~al.}(2012)\citenamefont {Singh},
  \citenamefont {Maignan}, \citenamefont {Pelloquin}, \citenamefont {Perez},\
  and\ \citenamefont {Simon}}]{Singh}%
  \BibitemOpen
  \bibfield  {author} {\bibinfo {author} {\bibfnamefont {K.}~\bibnamefont
  {Singh}}, \bibinfo {author} {\bibfnamefont {A.}~\bibnamefont {Maignan}},
  \bibinfo {author} {\bibfnamefont {D.}~\bibnamefont {Pelloquin}}, \bibinfo
  {author} {\bibfnamefont {O.}~\bibnamefont {Perez}}, \ and\ \bibinfo {author}
  {\bibfnamefont {C.}~\bibnamefont {Simon}},\ }\href {\doibase
  10.1039/C2JM16290C} {\bibfield  {journal} {\bibinfo  {journal} {J. Mater.
  Chem.}\ }\textbf {\bibinfo {volume} {22}},\ \bibinfo {pages} {6436} (\bibinfo
  {year} {2012})}\BibitemShut {NoStop}%
\end{thebibliography}
%

\end{document}